\documentclass[twocolumn,aps,pra,floatfix,showpacs,noshowkeys,epsfig,graphics]{revtex4-1}
\usepackage{graphicx}
\usepackage{amsmath}
\usepackage{amsfonts}
\usepackage{amssymb}

\newcommand{\newc}{\newcommand}
\newc{\beq}    {\begin{equation}}
\newc{\eeq}    {\end{equation}}

\newc{\beqa}    {\begin{eqnarray}}
\newc{\eeqa}    {\end{eqnarray}}
\newc{\bs}    {\section}
\newc{\no}    {\\ \nonumber}

\newc{\st}    {\stackrel}

\begin{document}
\title{ Quantum mechanics  emerges from information theory applied
to causal horizons}
\author{Jae-Weon Lee}\email{scikid@gmail.com}
\affiliation{ Department of energy resources development,
Jungwon
 University,  5 dongburi, Goesan-eup, Goesan-gun Chungbuk Korea
367-805}

\date{\today}

\begin{abstract}
It is suggested that  quantum mechanics
is not fundamental but emerges from classical information theory applied to  causal horizons.
The path integral quantization and quantum randomness can be
derived by considering information loss of  fields or particles
crossing Rindler horizons for accelerating observers.
 This implies that information is one of the fundamental roots
  of all physical phenomena.
  The connection between
this theory and Verlinde's entropic  gravity theory is also investigated.
 \end{abstract}

\maketitle

\section{introduction}

Quantum mechanics and relativity are two pillars of modern physics.
Recent developments of quantum information science revealed that
quantum mechanics and relativity miraculously cooperate so as not to violate each other.
For example, one can obtain a quantum mechanical state discrimination bound (Helstrom bound)
from the no-signaling condition of the special relativity ~\cite{PhysRevA.78.022335}.
This connection is very surprising, because a quantum mechanical phenomenon is stochastic while relativity is
deterministic.
Furthermore, the holographic principle~\cite{hooft-1993} or AdS/CFT correspondence~\cite{Maldacena}
also asserts an unexpected connection between classical gravity in bulk
and quantum mechanics on its surface. The origin of this mysterious connection is also still unknown.

Although quantum mechanics
is experimentally and mathematically well established, its
    origin is not identified.
  For example,  we do not know
 the origin of quantum randomness which leads to many paradoxes of quantum mechanics such
 as EPR paradox.
 This situation gave rise to numerous
 interpretations of quantum mechanics from the Copenhagen interpretation
 to the many worlds interpretation. One of the viable interpretations
 is information theoretic interpretation ~\cite{zeilinger1999,brukner-1999-83}, in which
 information about physical events plays a central role.
 For example, Zeilinger and Brukner ~\cite{zeilinger1999,brukner-1999-83}  introduced a notion of information space to
 describe quantum phenomena and suggested that quantum randomness aries from the discreteness of information (i.e., bits).

Why does physics have something to do with information? There are several hints for this question.
Landauer's principle in quantum information science
states that to erase  information of a system
irreversibly, energy should be consumed~\cite{landauer}. This means information is  physical.
In a series of works ~\cite{myDE,Kim:2007vx,Kim:2008re,Lee:2010bg,kias}, based on
this principle and quantum entanglement theory, the author and colleagues   suggested
that  information is the key to understand the origins of dark energy~\cite{myDE}, black hole mass~\cite{Kim:2007vx}
and even  Einstein gravity~\cite{Lee:2010bg,kias}.
For example, it is suggested that
a cosmic
causal horizon with a radius $R_h$ has
  temperature $T_h\sim 1/R_h\sim H$,   quantum informational entropy
$S_h\sim R_h^2$
and hence
a kind of  thermal
energy $E_{h}\sim T_h S_h\sim  M_P^2 R_h \sim  M_P^2 /H $
which is comparable to  dark energy observed~\cite{myDE}. Here, $M_P$ is the reduced Planck mass and $H$ is
 Hubble parameter. On the other hand, Jacobson~\cite{Jacobson} wrote a seminar paper linking the
 Einstein gravity to thermodynamics
 at Rindler horizons using the first law of thermodynamics $dE_h=T_h dS_h$.
 Recently,
  Verlinde~\cite{Verlinde:2010hp} (See also ~\cite{Padmanabhan:2009kr}) suggested
  fascinating  ideas linking gravitational force to entropic force
and derived  Newton's second law and the Einstein equation using similar horizon energy.
This brought many related studies
~\cite{Zhao:2010qw,Cai:2010sz,Cai:2010hk,Myung:2010jv,Liu:2010na,Tian:2010uy,Pesci:2010un,Diego:2010ju,
Vancea:2010vf,Konoplya:2010ak,Culetu:2010ua,Zhao:2010vt,Ghosh:2010hz,Munkhammar:2010rg,Kuang:2010gs,
Wei:2010am,Mureika:2010wk,Cai:2010zw,Morozov:2010ur,Konoplya:2010ak,Casadio:2010fs,Padmanabhan:2010xh,Horvat:2010wg,Pan:2010eu,Modesto:2010rm}.
In~\cite{Lee:2010xv} Lee suggested that Jacobson's~\cite{Jacobson}
gravity theory  or the quantum informational~\cite{Lee:2010bg}
theory can explain Verlinde's formalism of entropic gravity by
identifying the holographic
screen to be Rindler horizons for  accelerating observers. (See also ~\cite{Culetu:2010ce}).

Considering all these recent developments, it is plausible that quantum mechanics and gravity has information
as a common ingredient,
and information is the key to explain the strange connection between two.
If gravity and Newton mechanics can be derived by considering information
at Rindler horizons, it is natural
 to think quantum mechanics might have a similar origin.
In this paper, along this line, it is suggested that quantum field theory (QFT)
 and quantum mechanics
can be obtained from information theory
applied to causal (Rindler) horizons, and that quantum randomness aries from information blocking by the horizons.

 In Sec. II the connection between QFT and information theory is
 suggested.
In Sec. III the connection between our theory and  Verlinde's theory  is
investigated.
Section IV contains discussions.

\section{Quantum field theory from information theory}

Basic assumptions in this paper are followings.
First, we assume that the speed of information propagation (i.e., the light velocity $c$) is finite
and, hence, there are causal horizons.
Second, information is a fundamental ingredient of physics. That is, physics
should reflect information possessed by observers.
Third, we assume the general principle of relativity ($not$ general relativity of Einstein)
stating that
all observers  are equivalent with respect to the formulation of the fundamental laws of physics~\cite{moller}.
Finally, we also assume some notions of
classical information theory,
spacetime, and its coordinate transformations.

Let us begin by considering an
accelerating observer $\Theta_R$ with acceleration $a$ in $x_1$ direction
 in a flat spacetime with coordinates $X=(t,x_1,x_2,x_3)$ (See Fig. 1).
The Rindler coordinates $\xi=(\eta,r,x_2,x_3)$ for the observer are defined with
\beq
\label{rindler}
t= r~ sinh (a \eta),~ x_1= r ~cosh (a \eta)
\eeq
 on the Rindler wedges.
There is an inertial observer $\Theta_M$ too.
Now, consider a field $\phi$
 crossing the Rindler horizon  at a point $P$ (actually, a two dimensional surface in $x_2-x_3$ direction)
  and entering the future wedge $F$. A configuration for
 $\phi(x,t)$ is not necessarily meant to be classical but to be a  function of spacetime carrying information.

\begin{figure}[tpbh]
\includegraphics[width=0.3\textwidth]{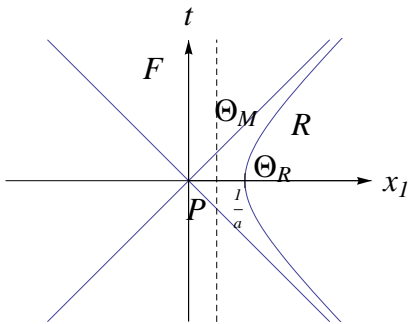}
\caption{ Rindler chart for  the observer $\Theta_R$ (curved line),
who has no accessible information about  field $\phi$
in  a causally disconnected  region $F$.
The observer can only estimate a probabilistic distribution of the field,
which turns out to be equal to that of a quantum field for inertial observer $\Theta_M$ (dashed line) in Minkowski spacetime.
}
\end{figure}

As the field enters  the Rinder horizon for the observer $\Theta_R$,
 the observer shall have no more information about  future configurations  of $\phi$
and  all what the observer can expect about $\phi$ evolution
 is a probabilistic distribution $P[\phi]$ of $\phi$ beyond the horizon.
Already known information about $\phi$ becomes  constraints for the distribution.
I suggest that this ignorance is the origin of quantum randomness.
According to our assumptions,
 information is fundamental in this theory, and  what determines the physics
in the wedge $F$
is not a deterministic classical physics  but the evolution of information itself.
  Thus, in this conjecture,
the physics in the wedge should reflect the ignorance of the observer $\Theta_R$ about the field
configurations, and there should not be a priori deterministic classical value for $\phi$.
That means for the observers there is no `objective physical reality' such as classical fields
before measurements. This is the main conjecture in this paper, which naturally follows if
we accept our assumptions.

This situation can be mathematically analyzed with classical information theory.
The maximum ignorance about the field can be expressed by maximizing the Shannon
information entropy $h[P[\phi]]$ of the possible (discrete) configurations
$\Phi=\{\phi_i(X)\}~, i=1\cdots n$ that the field may take beyond the horizon
with  probability $P[\phi_i]$.
A uniform probability distribution may be adequate when there is no
 information about the events represented by random variables $\Phi$.
However, if there is  a priori  information available represented by
 $l$ testable expectations  (not a quantum expectation yet)
\beq
\label{constraint}
\langle f_k\rangle\equiv  \sum_{i=1}^n P[\phi_i] f_k[\phi_i],~(k=1\cdots l),
\eeq
 we should use the principle of maximum entropy by Boltzmann to calculate the probability distribution
 $P[\Phi]$.
Here, $f_k,~(k=1\cdots l)$ is a function of $\Phi$  and $\langle f_k\rangle$ is its expectation value
with respect to $P[\Phi]$. According to the theorem,
 by maximizing the Shannon entropy
\beq
\label{hP}
h[P]=-\sum_{i=1}^n P[\phi_i] ln  P[\phi_i],
\eeq
with the constraints in Eq. (\ref{constraint})
one can obtain
 the following probability distribution
 \beq
 P[\phi_i]=\frac{1}{Z} exp\left[-\sum_{j=1}^l \lambda_j f_j(\phi_i)\right]
 \eeq
with a normalization constant (partition function)
$
Z=\sum_{i=1}^n  exp\left[-\sum_{j=1}^l\lambda_j f_j(\phi_i)\right],
$
where $\lambda_j$'s are Lagrangian multipliers satisfying the following relation
$
\langle f_k\rangle=- \frac{\partial}{\partial \lambda_k} ln Z.
$
Thus, the usual Maxwell-Boltzman distribution is a natural consequence of classical information theory, when
there is information loss with constraints.
Lisi ~\cite{lisi-2006} suggested  a related derivation of the partition function by
assuming  a `universal action reservoir'.

What constraints can we impose on the motion of the field crossing the horizon?
One constraint may come from the energy conservation
\beq
\sum_{i=1}^n  P[\phi_i]H(\phi_i) = E,
\eeq
where $H(\phi_i)$ is the Hamiltonian as a function of the field $\phi_i$ and $E$ is its expectation.
This comes from the fact that the energy expectation value of the field should not change.
Another trivial one is the unity of  the probabilities
$
\sum_{i=1}^n  P[\phi_i] = 1.
$
Then, from the above theorem, the probability estimated by the Rindler observer,
 subject to these constraints, should be
\beq
\label{P}
P[\phi_i] = \frac{1}{Z} \exp\left[- \beta H(\phi_i)\right],
\eeq
where $\beta$ is the Lagrangian multiplier.
Here, the partition function is
\beq
\label{Z}
 Z = \sum_{i=1}^n  \exp\left[- {\beta H(\phi_i)} \right]=tr~ e^{-\beta H},
 \eeq
 and the trace is assumed to be performed with a (classical) discrete vector basis.
 Below, we shall take a continuum limit.
It is important to recall that
  the field $\phi$ can not have a specific value before measurements according to our assumptions.
The classical field theory could be obtained  through extremization of $P[\phi_i]$ after establishing a QFT later.
What  is assumed here is that   for both of the  observers $\Theta_M$ and $\Theta_R$, $\phi$ could
have arbitrary values before measurements. $Z$ represents this uncertainty.

 As an example, let us  consider  a scalar field with Hamiltonian
 \beq
 H(\phi)=\int d^3 x \left[ \frac{1}{2}\left( \frac{\partial \phi}{\partial t}\right)^2+
 \frac{1}{2}\left( {\nabla \phi}\right)^2 + V(\phi) \right]
 \eeq
 with potential $V$.
 $H$ alone, without a guiding principle, does not fully give us dynamics, neither classical nor quantum.
  For the Rindler observer with the coordinates $(\eta,r,x_2,x_3)$  the proper time variance is
  $ard\eta$ and hence the Hamiltonian is changed to
  \beqa
 H_R =  \int_{r>0} dr dx_\bot~ ar
 [
  \frac{1}{2}\left( \frac{\partial \phi}{ar \partial \eta}\right)^2
 +   \no
 \frac{1}{2}
 \left(\frac{\partial \phi}{ \partial r} \right)^2+\frac{1}{2}
 \left( {\nabla_\bot \phi}\right)^2
 +  V(\phi)] ,
 \eeqa
where $\bot$ denotes the plane orthogonal to $(\eta,r)$ plane.
Then, Eq. (\ref{Z}) becomes Eq. (2.5) of Ref. ~\cite{1984PhRvD..29.1656U};
\beq
\label{Z_R}
 Z_R = tr~ e^{-\beta H_R}.
 \eeq
It is important to notice
  that $Z$ (and hence $Z_R$) here is not a quantum partition function but a statistical partition function yet.
$Z_R$ simply represents the statistical system corresponding to the uncertain
 field configurations  beyond the horizon.

Now, we need to show $Z_R$ is equivalent to a quantum partition function.
Fortunately, this is already done in Ref. ~\cite{1984PhRvD..29.1656U}.
Based on QFT, in the reference, it was  shown that the
real-time thermal Green's functions of the Rindler observer with $Z_R$ are equivalent to
the vacuum Green's function in Minkowski coordinates.
(See ~\cite{Crispino:2007eb}  for a review.)
Thus, as well known, the Minkowski vacuum is equivalent to thermal states for the Rindler observers.
What I have newly shown here is that the  thermal partition function $Z_R$ assumed in  Ref. ~\cite{Crispino:2007eb}
 is from information loss about  field configurations beyond the Rindler horizon
and, therefore, the QFT formalism is equivalent to  the purely information theoretic formalism suggested in this paper.
Recall that Eq. (\ref{Z_R}) was derived without using any quantum physics in this paper.
Since quantum mechanics can be thought to be single particle limit of QFT, this implies also
that quantum mechanics emerges from information theory applied to Rindler horizons and
is not fundamental. Although the reasoning is  simple, conclusions can be far-reaching.

To be concrete, it is worthy to briefly repeat the calculation  in Ref.~\cite{Crispino:2007eb}.
There, it was shown by analytical continuation
that in the Rindler coordinates $Z_R$ is $mathematically$ equivalent to
 \beqa
 Z_R &=&\no
   & &N_0\int_{\phi(0)=\phi(\beta')}
  D\phi ~exp\{-\alpha\int_0^{\beta'} d\tilde{\eta} \int_{r>0} dr dx_\bot~ ar \no
 & &  \left[
  \frac{1}{2}\left( \frac{\partial \phi}{ar \partial \tilde{\eta}}\right)^2
 +  \frac{1}{2}
 \left(\frac{\partial \phi}{ \partial r} \right)^2+\frac{1}{2}
 \left( {\nabla_\bot \phi}\right)^2
 +  V(\phi)\right]\},
 \eeqa
 where we explicitly denoted a constant  $\alpha$  having a dimension
 of  $1/H_Rt$ for a dimensional reason. Thus, $\beta=\alpha \beta'$.

The trace turned
   into the  periodic boundary condition $\phi(\tilde{\eta}=0)=\phi(\tilde{\eta}=\beta')$  as usual.
  By further  changing integration variables as
$\tilde{r}=r cos(a\tilde{\eta}),  \tilde{t}= rsin(a\tilde{\eta})$ and choosing $\beta'=2\pi/a\equiv 1/\alpha k_B T_U$
the  region of integration is transformed from $0\le \tilde{\eta} \le \beta', 0\le r \le \infty$
into the full two dimensional flat space $-\infty\le \tilde{t} \le \infty, -\infty\le \tilde{r} \le \infty$.
Of course, this specific  $\beta'$ value leads to Unruh temperature
$ T_U={ a}/{2\alpha \pi k_B }$, where $k_B$ is the Boltzman constant.
From the well-known QFT result,  one can find $1/\alpha$ to be  $\hbar$.
Since  $\hbar=1/\alpha$ is from the lagrange multiplier $\beta$,
the Planck constant $\hbar$ is associated with  the
change of $Z_R$ by energy change, that is, $\hbar$ is some fundamental temperature given by nature.

Then, the partition function becomes
\beqa
 Z^E_Q &=&  N_1\int
  D\phi ~exp\{- \alpha\int d\tilde{r}d\tilde{t} dx_\bot~  [
  \frac{1}{2}\left( \frac{\partial \phi}{ \partial \tilde{t}}\right)^2 \no
 &+& \frac{1}{2}
 \left(\frac{\partial \phi}{ \partial \tilde{r}} \right)^2+\frac{1}{2}
 \left( {\nabla_\bot \phi}\right)^2
 +  V(\phi)]\}\no
 &=&N_1\int
  D\phi ~exp\left\{-  \frac{I_E}{\hbar}\right\}.
 \eeqa

where $I_E$ is the Euclidean action for the scalar field in the inertial frame.
By analytic continuation $\tilde{t}\rightarrow it$,
one can see $Z^E_Q$ becomes  the usual zero temperature quantum mechanical partition function $Z_Q$ for $\phi$.
Since both of $Z_R$ and $Z_Q$ can be obtained from $Z^E_Q$ by analytic continuation, they are
physically equivalent as pointed out in Ref. ~\cite{Crispino:2007eb}.
A partition function contains all information about a statistical system.
 Thus, it is enough to show the equivalence of two partition functions
 to prove the equivalence of QFT and the information theoretic model
 suggested in this paper, once we accept the information theoretic origin of $Z_R$.
 Of course, one can reverse the logic and obtain $Z$ in Eq. (\ref{Z_R}) from $Z_Q$.
 Now we see that quantum fluctuations correspond to the ignorance of Rindler observers about the fields beyond Rindler horizons.

\section{Quantum mechanics and entropic gravity}

It is straightforward to extend the previous analysis to quantum mechanics for point particles.
We can imagine a point particle  at a point $P$
just crossing the Rindler horizon and entering the future wedge $F$.
Like the case of the   field,
the Rindler observer gets no more information from the particle.
This maximal ignorance is represented by probability distribution $P[x_i(t)]$
for the i-th possible path that the particle may take.

\begin{figure}[tpbh]
\includegraphics[width=0.4\textwidth]{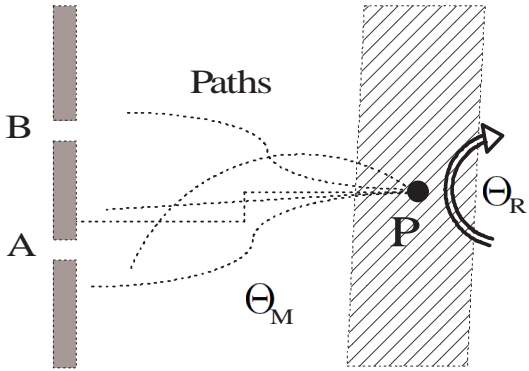}
\caption{
The Rindler observer $\Theta_R$ has no  more information about  paths of the particle
crossing  the horizon (shaded region)
and  all what the observer can expect about the particle is a probabilistic distribution
 of its motion. This seems to be the origin of quantum randomness of the motion.
 Here, $A$ and $B$ represent slits for a typical double-slit experiment.}
\end{figure}
Then, the partition function is
\beq
\label{ZR}
 Z_R = \sum_{i=1}^n  \exp\left[- {\beta H(x_i)} \right]=tr~ e^{-\beta H},
 \eeq
 where $H$ is the point particle Hamiltonian now.
 Since the usual point particle quantum mechanics
 is a non-relativistic and single particle limit of the quantum field theory,
 we expect $Z_R$ is equal to the quantum partition function for the particle
 with mass $m$  in Minkowski spacetime
 \beqa
 Z_Q &=& \no
 & & N_2\int  Dx \exp\left[-\frac{i}{\hbar}\int d\tilde{t}  \left\{\frac{m}{2}
 \left( \frac{\partial x}{\partial \tilde{t}}\right)^2 -  V(x)\right\}  \right]\no
 &=&N_1\int
  Dx ~exp\left\{- \frac{i}{\hbar} I(x_i)\right\},
\eeqa
 where $I$ is the action. Then,
as is well known one can associate each path $x_i$ with a wave function $\psi\sim e^{-iI}$,
which leads to  Schr\"{o}dinger equation for $\psi$~\cite{derbes:881}.

 This interpretation could shed a new light on the paradoxical behaviors of quantum particles.
For example, consider the double slit experiment.
 the Rindler observer $\Theta_R$, having no access to the information about  paths of the particle,
could not say which path of two slits (A or B) in Fig. 2 was chosen by the particle.
Otherwise, it will violate the no-signaling principle.
According to our conjecture, physics in the wedge $F$ should reflect this ignorance. Thus,
the particle could not have a deterministic path before measurement.
On the other hand, the observer $\Theta_M$ who can measure the paths, after he or she enters the horizon
 (For this observer light cones play a role of causal horizons.),
has a chance to know the ``which-way" information. This could induce the ``wave function collapse".
According to our theory,  wave functions or states are neither particles nor physical waves but just probability
functions
 about information, thus there is no need for a concern regarding
immediate superluminal changes of wavefuncions.

This theory also gives some new insights on the origin of Verlinde's  entropic gravity theory.
In his papers identities of information and its entropy are not so clearly given.
Therefore, there are several concerns ~\cite{Li:2010cj,Culetu:2010ua,Gao:2010yy,Lee:2010za,Culetu:2010ce} on
the  assumptions Verlinde took.
Two important concerns are  about the origin of
the  entropy variation formula (Eq. (\ref{dS}) below)  and identity of the holographic screen.

According to our theory the entropy which is associated with the entropic force is  the entropy $h[P(x_i)]$
about unobservable paths, estimated
by the Rindler observers.
Culetu~\cite{Culetu:2010ua} pointed out that if
the screen plays the role of a local Rindler horizon at $r=c^2/a$, Verlinde's entropy formula can be explained.
This interpretation is in accordance with Lee et al's proposal ~\cite{Lee:2010bg,Lee:2010xv}
that  the Einstein equation represents information erasing process
at Rindler horizons.

The results in this paper enhance these interpretations.
For the observer  at $r=c^2/a$, the Rindler Hamiltonian becomes a physical Hamiltonian
generating $\eta$ translation~\cite{Crispino:2007eb}. Thus, this distance $r$
is special.
The observer shall have no more path information of the non-relativistic
particle with mass $m$ crossing the horizon.
In this case
the  loss of information of the particle results in
the horizon entropy $S_h$ increase~\cite{Lee:2010xv} as
\beq
\label{dS}
 \Delta S_h=\frac{\Delta E_h}{T_U}=\frac{2\pi c k_B m r}{\hbar},
\eeq
which is just the entropy variation Verlinde assumed.
Here, we used  the Unruh temperature and the horizon energy variation $\Delta E_h\simeq mc^2$
due to the holographic principle.
Once we obtain this formula, it is straightforward to reproduce Newton's equation
and gravity in Verlinde's formalism with the equipartition energy law.

The maximum entropy proposal in Verlinde's theory can be also understood in this way.
From Eq. (\ref{ZR}) free energy can be expressed as
\beq
F=-\frac{1}{\beta} ln Z_R.
\eeq
The classical path corresponds to the saddle point approximation ($Z_R\sim exp[-\beta I_E(x_{cl})$] )~\cite{Banerjee:2010yd}
\beq
F\simeq F_{cl}= -\frac{1}{\beta} (-\beta I_E(x_{cl}))=I_E(x_{cl}),
\eeq
where $I_E(x_{cl})$ is the Euclidean action for classical path satisfying the Lagrange equation.
Since the maximum entropy is achieved when $F$ is minimized, we see that
classical physics with the minimum action corresponds to a maximum entropy (of the paths) condition, that is,
the classical path is the typical path  maximizing the entropy $h[P]$ with the constraints for the Rindler observer.
The holographic principle demands that the increase of the horizon entropy should be equal to
the entropy of the paths of the particle entered.

In short, Verlinde's holographic screen is just Rindler horizons and its entropy is
associated with the lost path information of the particle crossing  the horizons~\cite{Lee:2010xv}.
Then, there is an entropic force  linked to this information loss  which can be calculated
by using the first law or Landauer's principle.
Thus, our theory provides a natural  model for the entropy and information
which Verlinde assumed.

\section{Discussions}

In this paper it is shown that if there is a causal boundary
for an observer, the observer could expect statistical distribution  for physical objects beyond the horizon
due to information loss.
For  another observer who can access the objects this thermal distribution corresponds to just quantum fluctuation.

What are the merits of our   new interpretation of quantum physics?
First, this theory explains the strange connection between quantum mechanics and special relativity such
as no-signaling condition in quantum measurements.
Since our formalism of  quantum mechanics itself emerges from the limitation of the information propagation velocity,
 it is natural that we can not send a classical superluminal signal even with
quantum nonlocal correlation (entanglement) by any means.
Second, from this fact, it might give us a new hint to  study of unification of gravity and quantum mechanics.
Third, this model could also explain the origin of Verlinde's formalism about Newton mechanics and gravity.
Our theory is in accordance  with the quantum informational dark energy model~\cite{myDE} too.

In summary,
it is shown that the path integral quantization and quantum randomness can be
derived by considering information loss behind Rindler horizons.
Quantum mechanics is not fundamental and emerges from information theory accompanied with
the Rindler coordinate transformation.
This implies that quantum mechanics is  more about information
rather than particles or waves.
Thus, now we have some striking  relationships among information, gravity, Newtonian mechanics, and even quantum mechanics.
Information seems to be one of the roots of all physical phenomena.

\section*{acknowledgments}
This work was supported in part by Basic Science Research Program through the
National Research Foundation of Korea (NRF) funded by the ministry of Education, Science and Technology
(2010-0024761) and the topical research program (2010-T-1) of Asia Pacific Center for Theoretical
Physics.
%

\end{document}